\documentstyle[preprint,revtex]{aps}
\begin{document}
\preprint{LA-UR-92-3726}
\widetext
\begin{title}
Intrinsic Parity of the $(j,0)\oplus(0,j)$ Mesons\footnotemark[1]
\footnotetext[1]{This work was done under the auspices of the U. S.
Department of Energy.}
\end {title}
\author{D. V. Ahluwalia}
\begin{instit} Medium Energy Physics Division, LAMPF,  MS H-844, Los Alamos
National Laboratory\\ Los Alamos, New Mexico 87544, USA; E-mail:
av@lampf.lanl.gov \end{instit}
\begin{abstract}
Contrary to the usual belief,  by carefully examining the
operation of parity transformation on the $(1,0)\oplus(0,1)$ mesons in the
generalized canonical representation,  we establish that the $(j,0)\oplus(0,j)$
meson-antimeson pair have {\it opposite} intrinsic parity. This opens up the
possibility that while the  particles without an internal structure may utilize
one representation of the Lorentz group, phenomenologies of composite particles
may exploit a different representation. As such (perhaps, only some of)   the
meson structures beyond the standard $q\,\overline {Q}$ may exploit the
$(j,0)\oplus(0,j)$ generalized canonical representation of the Lorentz group --
this would result in a meson and the associated antimeson to manifest
themselves in {\it different} partial waves.
\end{abstract}


\newpage

In this letter we argue that the relative intrinsic-parity\footnotemark[2]
\footnotetext[2]{There is a slight ambiguity even in the definition of {\it
relative} intrinsic parity. For this ambiguity the curious reader is referred
to footnote [11] on p. 570 of Ref. \cite{Weis}.} of the particle-antiparticle
pair emerges as a (Lorentz group) representation-dependent kinematical object,
in apparent contradiction to the usual\footnotemark[3]\footnotetext[3]{
For example, see Ref. [1, footnote 13] and Ref. \cite{WeinbergA}.} belief.
Our
motivation for this investigation is to continue our {\it ab initio}  study,
 based on  Refs. \cite{Weinberg,Wigner}, to establish a
relativistic phenomenology \cite{High} for high spin hadronic resonances.
These resonances
will become increasingly more accessible at CEBAF, NIKHEF, RHIC, a possible
upgrade of LAMPF, and other new medium energy nuclear physics facilities.

For concreteness we look at the relativistic phenomenology of the $j=1$ mesons.
Following conventions defined by Ryder \cite{Ryder} the $(1/2,\,1/2)$
representation of the Lorentz group corresponds to the description of the $j=1$
matter fields in terms of the vector field $A^\mu(x)$. The vector field
$A^\mu(x)$ satisfies the Proca equation. Within this framework the relative
intrinsic parity of the meson-antimeson pair is {\it same}.
By considering the $(1,0)\oplus(0,1)$ representation
in detail we will show that the  relative intrinsic parity for the
meson-antimeson, described by the generalized canonical [4a,b] representation
$(1,0)\oplus(0,1)$ matter field, is {\it opposite. } The generalization of this
result to the mesons with $j > 1$ will be seen as essentially obvious. As noted
in the abstract, this result opens up the possibility that while structureless
fundamental particles may utilize one representation of the Lorentz group, say
(1/2,\,1/2), the composite particles may find their phenomenological
description
in terms of other representations, such as $(1,0)\oplus(0,1)$.

We begin with the classical considerations similar to the
ones found for the $(1/2,0)\oplus(0,1/2)$ Dirac field in the
standard texts, such as Nachtmann's treatment in Ref. [6, Sec. 4.5].
The $(1,0)\oplus(0,1)$ wave function
satisfies the spin one Weinberg [1, 4h] equation
\begin{equation}
\left(\gamma_{\mu\nu}\,\partial^\mu\,\partial^\nu \,+\,m^2\right)\,
\psi(t,\,\vec x\,)\,=\,0\quad.
\label{one}
\end{equation}
The $6\times 6$ ten $\gamma_{\mu\nu}$ matrices in the generalized canonical
representation [4h] are given by
\begin{eqnarray}
\gamma_{\circ\circ}\,=\,
\left(
\begin{array}{ccc}
I &{\,\,} &0\\
0 &{\,\,} &-I
\end{array}
\right),\quad
\gamma_{\ell\circ}\,=\,\gamma_{\circ\ell}\,=\,
\left(
\begin{array}{ccc}
0 &{\,\,}& -J_\ell\\
J_\ell&{\,\,} & 0
\end{array}
\right), \nonumber \\ \nonumber \\
\gamma_{\ell\jmath}\,=\,\gamma_{\jmath\ell}\,=\,
\left(
\begin{array}{ccc}
I & {\,\,}&0\\
0 &{\,\,}&-I
\end{array}
\right)
\,g_{\ell\jmath} \,+\,
\left(
\begin{array}{cc}
\{J_\ell,J_\jmath\} & 0\\
0 & -\,\{J_\ell,J_\jmath\}
\end{array}
\right)\quad;
\label{two}
\end{eqnarray}
where $\vec J$ are the $3\times 3$ spin one matrices with $J_z$ diagonal;
 $I$ are the $3\times 3$ identity matrices; $g_{\mu\nu}$ is the
flat spacetime metric with $diag\,(1,\,-1,\,-1\,-1)$; $\ell,\,\jmath$
run over the spacial indices $1, \,2, \,3$; and $\{J_l,J_\jmath\}$ is the
anticommutator of $J_\ell$ and $J_\jmath$ .
We seek the parity-transformed wave function\footnotemark[4]
\footnotetext[4]{The operation of
parity is defined via $x'^\mu\,=\,{\Lambda^\mu}_\nu\, x^\nu$, with
$\Lambda_P\,\equiv\, [{\Lambda^\mu}_\nu]\,=\,diag (1,\,-1,\,-1,\,-1)$. All
conventions, unless indicated otherwise, are that of Ref. \cite{Ryder}. }
\begin{equation}
\psi'(t',\,{\vec x}\,') \,=\,S(\Lambda_P)\,\psi(t,\,\vec x)\quad,
\label{three}
\end{equation}
such that Eq. (\ref{one}) holds true for $\psi(t',\,{\vec x}\,')$
\begin{equation}
\left(\gamma_{\mu\nu}\,\partial^{\,\prime\mu}\,\partial^{\,\prime\nu}
 \,+\,m^2\right)\,
\psi(t',\,{\vec x}\,'\,)\,=\,0\quad.
\label{four}
\end{equation}
It is a straight forward exercise to find that $S(\Lambda_P)$ must
simultaneously satisfy the following requirements
\begin{eqnarray}
S^{-1}(\Lambda_P)\,\gamma_{\circ\circ}\,S(\Lambda_P)\,=\,\gamma_{\circ\circ},
\quad
S^{-1}(\Lambda_P)\,\gamma_{\circ\jmath}\,S(\Lambda_P)\,=\,-\,
\gamma_{\circ\jmath}\quad,
\nonumber\\
S^{-1}(\Lambda_P)\,\gamma_{\jmath\circ}\,S(\Lambda_P)\,=\,-\,
\gamma_{\jmath\circ},
\quad
S^{-1}(\Lambda_P)\,\gamma_{\ell\jmath}\,S(\Lambda_P)\,=\,\gamma_{\ell\jmath}
\quad.
\end{eqnarray}
Referring to Eqs. (\ref{two}), we now note that while $\gamma_{\circ\circ}$
commutes  with $\gamma_{\ell \jmath}$ it
anticommutes with $\gamma_{\circ\jmath}$
\begin{equation}
\left[\gamma_{\circ\circ},\,\gamma_{\ell\jmath}\right]\,=\,
\left[\gamma_{\circ\circ},\,\gamma_{\jmath\ell}\right]\,=\,0,\quad
\left\{\gamma_{\circ\circ},\,\gamma_{\circ\jmath}\right\}\,=\,
\left\{\gamma_{\circ\circ},\,\gamma_{\jmath\circ}\right\}\,=\,0\quad.
\end{equation}
As a result, confining to the norm preserving transformations (and ignoring a
possible {\it global} phase factor), we  identify $S(\Lambda_P)$ with
$\gamma_{\circ\circ}$, yielding
\begin{equation}
\psi'(t',\,{\vec x}\,') \,=\,\gamma_{\circ\circ}\,\psi(t,\,\vec x)\quad
\Longleftrightarrow\quad
\psi'(t',\,{\vec x}\,') \,=\,\gamma_{\circ\circ}\,\psi(t',\,-\,{\vec x}\,')
\quad,\label{seven}
\end{equation}

This prepares us to proceed to the field theoretic considerations.
The $(1,0)\oplus(0,1)$ matter field operator may be defined as follows
[4a]

\begin{eqnarray}
\Psi(x) &\,= \,&
\sum_{\sigma\,=\,0,\,\pm 1}
\int {d^3p\over (2\pi)^{3} } {1\over 2\,\omega_{\vec p}} \nonumber\\
 &\quad\times&\Big[ u_\sigma(p)\, a_\sigma(\vec p\, )\, \exp(-i p\cdot x)
+  v_\sigma(p\,) \,b^\dagger_\sigma(\vec p\,) \,\exp(+i p\cdot x) \Bigr ]
\quad,\label{eight}
\end{eqnarray}
with $\omega_{\vec p\,} = \sqrt{m^2 + {\vec p\,}^2}$.
The explicit  general-canonical-representation
expressions  [4h,b,f] for the
$(1,0)\oplus(0,1)$ spinors $u_\sigma(\vec p\,)$ and $v_\sigma(\vec p\,)$,
which appear in Eq. (\ref{eight}), are
{\footnotesize
\begin{eqnarray}
u_{_ {+1}}( p)&\,=\,&
\pmatrix{m+\left[(2p_z^2~+~p_{_{+}} p_{_{-}}) / 2(E+m)\right]\cr\cr
                      {p_z p_{_{+}}/{\sqrt 2}(E+m)}\cr\cr
              { p_{_{+}}^2/ 2(E+m) }\cr\cr
               p_z\cr\cr
                   {p_{_{+}}/{\sqrt 2}}\cr\cr
                   0\cr}\quad,
u_{_{0}}( p)\,=\,\pmatrix{{p_z p_{_{-}}/{\sqrt 2}(E+m)}\cr\cr
                      m+\left[{p_{_{+}} p_{_{-}}/(E+m) }\right]\cr\cr
                       -{p_z p_{_{+}}/{\sqrt 2}(E+m)}\cr\cr
                       {p_{_{-}}/{\sqrt 2}}\cr\cr
                          0\cr\cr
                        {p_{_{+}}/{\sqrt 2}}\cr}\quad,\nonumber\\
u_{_{-1}}( p)&\,=\,&\pmatrix{ { p_{_{-}}^2/ 2(E+m) }\cr\cr
                             -{p_z p_{_{-}}/{\sqrt 2}(E+m)}\cr\cr
                   m+\left[{(2p_z^2~+~p_{_{+}} p_{_{-}})/ 2(E+m)}\right]\cr\cr
                      0\cr\cr
                      {p_{_{-}}/{\sqrt 2}}\cr\cr
                   -p_z\cr}\quad,\qquad
v_\sigma(p) \,=\,
\left(
\begin{array}{ccc}
0 & {\quad}I \\
I & {\quad}0
\end{array}\right)
\,u_\sigma(p)\label{nine}\quad.
\end{eqnarray}
}
In the above expression we have defined $p_\pm\,=\,p_x\,\pm\,i\,p_y$.
The transformation properties of  the ``particle'' [``antiparticle''] creation
operators $a^\dagger_\sigma(\vec p\,)\quad [b^\dagger_\sigma(\vec p\,)]$ are
obtained from the condition
\begin{equation}
U(\Lambda_P)\,\Psi(t',\,\vec x\,'\,)\,U^{-1}(\Lambda_P)\,=\,
\gamma_{\circ\circ}\,\Psi(t',\,-\,\vec x\,'\,)\label{ten}
\quad,
\end{equation}
where $U(\Lambda_P)$ represents a unitary operator which governs the operation
of parity in the Hilbert space of the single particle-antiparticle  states.
Using the definition of $\gamma_{\circ\circ}$, Eqs. (\ref{two}), and the
explicit expressions for the
$(1,0)\oplus(0,1)$ spinors $u_\sigma(\vec p\,)$ and $v_\sigma(\vec p\,)$
given by Eqs. (\ref{nine}), we find
\begin{eqnarray}
\gamma_{\circ\circ}\, u_\sigma(p')\,=\,+\,u_\sigma(p)\quad,\nonumber\\
\gamma_{\circ\circ}\, v_\sigma(p')\,=\,-\,v_\sigma(p)\quad,
\label{elev}
\end{eqnarray}
with $p'$  the parity-transformed $p$  --- i.e.
for $p^\mu\,=\,(p^\circ,\,\vec p\,)$,
$p^{\prime\mu}\,=\,(p^\circ,\,-\vec p\,)$. The observation (\ref{elev})
when coupled with the requirement (\ref{ten}) immediately yields the
transformation properties of the
particle-antiparticle creation operators
\begin{eqnarray}
U(\Lambda_P)\,a^\dagger_\sigma(\vec p\,)\, U^{-1}(\Lambda_P)
\,&=&\,+\,a^\dagger_\sigma(-\,\vec p\,)\nonumber\\
U(\Lambda_P)\,b^\dagger_\sigma(\vec p\,)\, U^{-1}(\Lambda_P)
\,&=&\,-\,b^\dagger_\sigma(-\,\vec p\,)\quad.
\label{twel}
\end{eqnarray}
Under the assumption that the vacuum is invariant under the parity
transformation, $U(\Lambda_P)\,\vert\quad\rangle\,=\,\vert\quad\rangle$, we
arrive at the result that the ``particles'' (described by the u-spinors) and
``antiparticles'' (described by the v-spinors) have opposite relative
intrinsic parities
\begin{eqnarray}
U(\Lambda_P)\,\vert\vec p,\,\sigma\rangle^u\,=\,+\,\vert\,-\vec p, \sigma
\rangle^u \quad, \nonumber\\
U(\Lambda_P)\,\vert\vec p,\,\sigma\rangle^v\,=\,-\,\vert\,-\vec p, \sigma
\rangle^v \quad.\label{thir} \end{eqnarray}

The results (\ref{twel}) and (\ref{thir}) are precisely what we set out to
prove. While the particle-antiparticle pairs in the $(1/2,\,1/2)$
representation of the Lorentz group have {\it same} relative intrinsic-parity,
the particle-antiparticle pairs in the $(1,0)\oplus(0,1)$ representation
have {\it opposite} intrinsic parity. Because of the general structure of the
Weinberg's equations and the $(j,0)\oplus(0,j)$ spinors,
we assert\footnotemark[5]\footnotetext[5]{The reader may wish to note that all
$(j,0)\oplus(0,j)$ spinors satisfy relations very similar to Eq. (\ref{elev});
etc.} that
this result is true for {\it all} spins. Whether (at least some of the)
mesons beyond the standard $q\,\overline Q$ structures
exploit the $(j,0)\oplus(0,j)$ representation of the Lorentz group remains an
open experimental question. {\it
To sum up, the analysis of this work establishes that the relative intrinsic
parity of a particle-antiparticle pair
is a Lorentz-representation-dependent kinematical object. Which of the
various representations, for a given spin, is actually realized in nature, such
as in constructing the phenomenologies of composite particles, can be
(contrary to the usual belief --- for example, see Ref. \cite{WeinbergA})
experimentally determined.}

\acknowledgements

An anonymous referee (of a related work) is to be thanked for bringing to my
attention footnote 13 of Ref. \cite{Weinberg}. Mikkel Johnson and Mikolaj
Sawicki are  to be thanked for being {\it insistent} that we understand
intrinsic-parity at a deeper level within the context of our other
collaborative efforts  --- in addition they kindly read the rough draft of this
work and provided comments and suggestions. It is also my pleasure to extend
thanks to Dick Arnowitt,  Terry Goldman and Barry Holstein for conversations on
the subject matter of this work. Finally, I thankfully acknowledge financial
support via a postdoctoral fellowship by the Los Alamos National Laboratory.

 \end{document}